\pdfoutput=1
\documentclass[a4paper,fleqn]{cas-dc}
\usepackage[authoryear]{natbib}

\usepackage{bm}
\usepackage{algorithm}
\usepackage{algpseudocode}
\newcommand{\vect}[1]{{\bm{\mathit{#1}}}}  
\newcommand{\matr}[1]{\mathbfsf{#1}}  
\hypersetup{citecolor=blue,linkcolor=blue,urlcolor=blue}

\newcommand{\T}{^\mathrm{T}}

\ExplSyntaxOn
\cs_set:Npn \__first_footerline:
{
  \group_begin:
  \small
  \sffamily
  \__short_authors: :~
  { \rmfamily \itshape Preprint }
  \group_end:
}
\ExplSyntaxOff

\begin{document}

\let\WriteBookmarks\relax
\renewcommand{\topfraction}{0.9}
\renewcommand{\dbltopfraction}{0.9}
\renewcommand{\textfraction}{0.07}
\renewcommand{\floatpagefraction}{0.8}
\renewcommand{\dblfloatpagefraction}{0.8}
\setcounter{topnumber}{3}
\setcounter{dbltopnumber}{2}

\shorttitle{Washout-free echo state networks}
\shortauthors{A. N\'ovoa}

\title[mode=title]{Washout-free synchronization, parameter estimation and uncertainty quantification of echo state networks}

\author[inst1,inst2]{Andrea N\'ovoa}[orcid=0000-0003-0597-8326]
\cormark[1]
\ead{a.novoa@imperial.ac.uk}

\affiliation[inst1]{organization={Aeronautics Department, Imperial College London},
            addressline={Exhibition Road},
            city={London},
            postcode={SW7 2AZ},
            country={UK}}

\affiliation[inst2]{organization={I-X AI in Science Centre, Imperial College London},
            addressline={White City Campus},
            city={London},
            postcode={W12 0BZ},
            country={UK}}
\cortext[1]{Corresponding author}

\begin{abstract}
Echo state networks (ESNs) forecast chaotic systems for several Lyapunov times. Like all recurrent networks, ESNs require a washout before forecasting, i.e., a phase in which the full state drives the network until the reservoir state synchronizes with the system. At deployment, however, sensors measure only part of the state, with noise and at a lower sampling rate than the training data.
We propose a washout-free deployment of ESNs via ensemble data assimilation. In closed loop, the ESN is a Markovian dynamical system; hence, a sequential filter can correct its reservoir state whenever data are available. The filter synchronizes an ensemble of reservoir states, drawn from the training data, with the system, and quantifies the forecast uncertainty. We extend the method to parametric ESNs, to infer physical parameters, and to parallel ESNs with localization, to forecast high-dimensional systems.
We test the method on the Lorenz-63 and Lorenz-96 systems. First, the filter synchronizes the ESN with a fraction of the data of a washout. Second, the filter deploys one full-state ESN from a subset of its components, which a washout cannot do. Third, the filter estimates a forcing that is not in the training data, and the estimated forcing lengthens the forecast. Fourth, localization synchronizes a parallel ESN with eight times fewer members than the global filter. Fifth, the ensemble spread signals, without the true state, when the ESN synchronizes and when the filter fails. This work enables the deployment of ESNs from partial, sparse and noisy observations.
\end{abstract}


\maketitle

\section{Introduction}
\label{sec:intro}

Echo state networks (ESNs) are recurrent neural networks in which a fixed, random reservoir expands the input history nonlinearly, and only a linear readout is trained via linear regression~\citep{maass2002real,jaeger2004harnessing}. 
Despite this simple training, ESNs can approximate any fading-memory filter, i.e., any causal map from uniformly bounded inputs whose output depends less and less on older inputs~\citep{GRIGORYEVA2018495}, and forecast chaotic and turbulent flows for several Lyapunov times~\citep{vlachas_backpropagation_2020,doan2021short,racca_data_2022}.  
Two extensions make ESNs suitable for high-dimensional and parameter-dependent systems. First, in parallel ESNs~\citep{pathak_model_2018}, each of several small reservoirs forecasts a local region of the domain, which scales ESNs to high-dimensional systems without encoding the state onto a low-dimensional space, e.g., with convolutional autoencoders~\citep{racca_predicting_2023}. Second, in parametric ESNs, a physical parameter is an additional input of the reservoir~\citep{kong_machine_2021}, so that one ESN forecasts the system over a range of parameters.

At deployment, a trained ESN must first be synchronized with the measured system, because the reservoir state at the start of the forecast is unknown. The synchronization is performed in the \emph{washout}  phase, in which the measured state is the input at every time step until the reservoir forgets its initial condition. 
By the echo state property, the reservoir state depends on the input history only~\citep{jaeger2004harnessing}. 
The washout, therefore, requires (i) measurements of all the state components on which the ESN is trained, and (ii) measurements at every time step.
Other recurrent neural networks, such as long short-term memory networks and gated recurrent units, also start from an unknown hidden state, and require the same phase, known in machine learning as warm-up, and therefore the same data~\citep{vlachas_backpropagation_2020}.
In many applications, neither requirement is met: sensors measure only some components of the state, at a lower sampling rate than the training data, for example, in fluid mechanics~\citep{novoa_magri_2022}, weather forecasting~\citep{carrassi2018data} and structural health monitoring~\citep{farrar2007introduction}.
{A common remedy is to train the ESN on the measured components only, and to overwrite its input with the data whenever they arrive, which is known as direct replacement~\citep{lu_reservoir_2017,ozalp_reconstruction_2023,carlson2026nudging}. This remedy requires one ESN for each set of sensors, degrades as fewer components are measured, and provides no estimate of uncertainty.}

Data assimilation combines a model with observations to estimate the state, and possibly the parameters, of a system, together with their uncertainty~\citep{sanz2023inverse,evensen_data_2009}. 
In sequential data assimilation, the model forecasts the state until data are available, and a Bayesian update (the analysis) corrects the forecast with the data, each weighted by its uncertainty. 
In ensemble filters, the probability distribution of the state is approximated by a Monte Carlo sample, i.e., an ensemble of model states, whose sample covariance approximates the uncertainty. Because the number of ensemble members is typically much smaller than the number of state variables, the sample covariance contains spurious long-range correlations, which localization removes by restricting each update to nearby observations~\citep{hamill2001distance,hunt2007efficient}.  Data assimilation is the backbone of numerical weather prediction~\citep{houtekamer2016review} and requires neither the full state nor data at every time step, which are the two requirements of the washout.

Data assimilation and machine learning have mostly been combined during training, when data assimilation provides the states from which a reduced-order model or a model-error correction is learned~\citep{bocquet_bayesian_2020,brajard_combining_2020,wikner_using_2021}. After deployment, data assimilation can also provide the data with which a model is retrained~\citep{zighed_efficient_2026}, with which the readout of an ESN is fine-tuned~\citep{novoa2025}, or the state estimates from which a reinforcement-learning agent controls a chaotic system~\citep{ozan2025data}. 
Kalman filters have also been employed to train the weights of recurrent neural networks~\citep{haykin2004kalman_ch2}. In these approaches, as in deep ensembles and Monte Carlo dropout~\citep{gal_dropout_2016}, the uncertain quantities to estimate are the weights of the network. In an ESN, however, the weights are fixed after training, and the uncertainty at deployment lies in the reservoir state, which is unknown until the ESN is synchronized. \citet{penny_integrating_2022} employed trained ESNs as the forecast model of ensemble and variational data assimilation, but initialized the ensemble by a washout of the full state, which requires data that are not available at deployment.

In this work, we recast the deployment of a trained ESN as sequential Bayesian state estimation. An ensemble of reservoir states, initialized from states visited during training, is forecast in closed loop and is updated at every observation by an ensemble square-root Kalman filter~\citep{evensen_ensemble_2003,sakov_implications_2008}. We show how different data assimilation tools can be used to 
The contributions are to
(i) synchronize the ESN without a washout, and compare the data that the filter and a washout require (\S\ref{sec:sync});
(ii) deploy an ESN trained on the full state from a subset of its components (\S\ref{sec:partial_results});
(iii) estimate an unknown physical parameter, which the parametric ESN reads in its readout only, together with the reservoir state (\S\ref{sec:param_results});
(iv) localize the analysis to synchronize a parallel ESN of a high-dimensional system with few members (\S\ref{sec:loc_results}); and
(v) show that the ensemble spread quantifies the uncertainty of the state and of the forecast, and signals a failed synchronization without the true state (\S\ref{sec:uq_results}).

\section{Echo state networks}\label{sec:method}
Consider a dynamical system whose true state, $\vect{u}^\dagger(t_n) \in \mathbb{R}^{N_u}$, is sampled at discrete times, $t_n$. 
An ESN approximates the dynamics of $\vect{u}^\dagger$ in three steps~\citep{jaeger2004harnessing,lukovsevivcius_practical_2012}. 
First, the ESN maps its input onto a reservoir state, $\vect{r}(t_n) \in \mathbb{R}^{N_r}$, with $N_r \gg N_u$. 
Second, the reservoir state evolves in time with a nonlinear map
\begin{subequations}\label{eq:esn}
\begin{align}
    \vect{r}(t_{n+1}) = \tanh\!\left(\matr{W}\vect{r}(t_n)
               + \matr{W}_{\mathrm{in}}\,\vect{u}_{\mathrm{in}}(t_n) + \vect{b}_{\mathrm{in}}\right),
    \label{eq:reservoir}
\end{align}
where $\vect{u}_{\mathrm{in}} \in \mathbb{R}^{N_\mathrm{in}}$ is the input; 
$\vect{b}_{\mathrm{in}}$ is the input bias; 
and $\matr{W}_{\mathrm{in}} \in \mathbb{R}^{N_r \times N_\mathrm{in}}$ and $\matr{W} \in \mathbb{R}^{N_r \times N_r}$ {are the input and reservoir matrices, which are random and fixed}.
Each row of $\matr{W}_{\mathrm{in}}$ has one non-zero entry, i.e., each unit reads one input component, with a weight sampled uniformly from $[-\sigma_{\mathrm{in}}, \sigma_{\mathrm{in}}]$, where $\sigma_{\mathrm{in}}$ is the input scaling; 
and $\matr{W}$ is scaled to a spectral radius, $\rho_\mathrm{s}$. 
Third, the ESN projects the reservoir state back onto the physical state with a linear readout
\begin{align}
    \vect{u}(t_{n+1}) = \matr{W}_{\mathrm{out}}\, [\vect{r}(t_{n+1}); 1],
    \label{eq:esn_update}
\end{align}
\end{subequations}
where the readout matrix, $\matr{W}_{\mathrm{out}} \in \mathbb{R}^{N_u \times (N_r+1)}$, is the only part of the ESN that is trained. 
To train it, we drive the ESN with a training record of $T_\mathrm{train}$ steps. 
The first $T_w$ steps are a washout, which makes the reservoir state independent of its arbitrary initial condition (\S\ref{sec:intro}). 
After the washout, we collect the augmented reservoir states{, $[\vect{r};1]$,} in the columns of $\tilde{\matr{R}} \in \mathbb{R}^{(N_r+1) \times (T_\mathrm{train}-T_w)}$, and the targets, i.e., the full state one step later, in the columns of $\hat{\matr{Y}} \in \mathbb{R}^{N_u \times (T_\mathrm{train}-T_w)}$. 
Because Eq.~\eqref{eq:esn_update} is linear in $\matr{W}_{\mathrm{out}}$, the readout is the solution of a ridge regression
\begin{equation}
    \matr{W}_{\mathrm{out}} = \hat{\matr{Y}}\,\tilde{\matr{R}}\T
            \!\left(\tilde{\matr{R}}\,\tilde{\matr{R}}\T
            + \beta\,\mathbb{I}\right)^{-1},
    \label{eq:ridge}
\end{equation}
where $\beta$ is the Tikhonov regularization factor. 
The ESN runs in open loop or in closed loop. 
In closed loop, the ESN feeds its output back as its input, and Eq.~\eqref{eq:esn} becomes an autonomous map, which forecasts the system. 
Before the closed-loop forecast, the ESN must be synchronized with the system, which is done in the washout phase (\S\ref{sec:intro})
in open loop (teacher forcing). In open loop, the input is observed data. 
\begin{figure}
  \centering
  \includegraphics[width=\columnwidth]{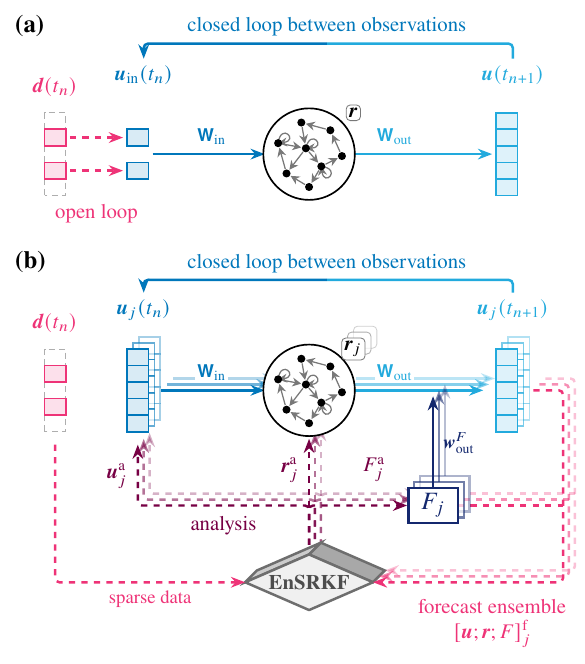}
  \caption{Deployment of an ESN from partial observations. (a) Direct replacement in the partial-input ESN (\S\ref{sec:direct_method}): the observed data replace the measured inputs in open loop. (b) Data assimilation (\S\ref{sec:analysis}): an ESN forecasts independently  $m$ ensemble members (stacked boxes). When observations are available, the EnSRKF updates the ensemble of reservoir states and the physical state, as well as the ensemble of parameters on the parametric ESN  (\S\ref{sec:param_method}).}
  \label{fig:schematic}
\end{figure}

\subsection{Partial-input ESN}\label{sec:partial_method}

In practice, only $N_q < N_u$ components of the state are measured, which we collect in the observed state, $\vect{q}^\dagger(t_n) \in \mathbb{R}^{N_q}$. 
The system is observed through the noisy data
\begin{equation}
    \vect{d}(t_n) = \vect{q}^\dagger(t_n) + \vect{\epsilon}(t_n), \quad
    \vect{\epsilon}(t_n) \sim \mathcal{N}(\vect{0},\,\matr{C}_{\epsilon\epsilon}),
    \label{eq:obs_model}
\end{equation}
where $\matr{C}_{\epsilon\epsilon}$ is the covariance of the observation noise. The data are available only at the observation times.   
A full-state ESN reads the full state, $\vect{u}_{\mathrm{in}} = \vect{u}$ with $N_\mathrm{in} = N_u$. 
Its washout requires a contiguous record of the full state, which is not available in practice. 
Alternatively,  a partial-input ESN reads the measured components, $\vect{u}_{\mathrm{in}} = \vect{q}$ with $N_\mathrm{in} = N_q$, and returns the full state~\citep{lu_reservoir_2017,ozalp_reconstruction_2023}. 
The partial-input ESN can be washed out with the measured components alone, and in closed loop it reads the measured components of its output, as depicted in Fig.~\ref{fig:schematic}a.

\subsection{Deployment by direct replacement}\label{sec:direct_method}

At deployment, the forecast of an ESN (either full-state or partial-input) can be corrected by overwriting the measured components of its input with the data at each observation time{, $\vect{u}_{\mathrm{in}}(\vect{x}_d) \leftarrow \vect{d}$, where $\vect{x}_d$ are the indices of the measured components}.
This correction is known as direct replacement, and it is the limit of nudging (or continuous data assimilation) in which the measured components relax to the data in one step, i.e., with unit gain~\citep{carlson2026nudging}.
Between observations{,} the ESN runs in closed loop (Fig.~\ref{fig:schematic}a). 
Direct replacement (i) degrades as fewer components are measured, because the ESN reconstructs the unmeasured components from the history of the measured components; 
(ii) with the partial-input ESN, requires one ESN for each set of sensors; 
and (iii) provides no estimate of uncertainty{.}

\section{Washout-free ESN deployment}\label{sec:analysis}

We deploy one full-state ESN from partial observations by performing state estimation via ensemble data assimilation. In closed loop, the ESN is an autonomous map of the input and the reservoir state (\S\ref{sec:method}), i.e., the next state depends on the current state only (the Markov property). The filter, therefore, needs no record of past inputs: it corrects the current state with the current data. Instead of driving one reservoir state with a contiguous record, we draw an ensemble of $m$ reservoir states, $\vect{r}_j$ with $j = 1, \ldots, m$, {from washouts on random windows of held-out training data, which consume no data of the deployed system} (\S\ref{sec:experiments}). The mean and covariance of the ensemble approximate the expected state and its uncertainty~\citep{evensen_data_2009}. Between observations, the ESN forecasts every member in closed loop with the same weights. At each observation, the filter corrects every member with the data, as depicted in Fig.~\ref{fig:schematic}b. 
Unlike the washout and direct replacement (\S\ref{sec:direct_method}), the filter assimilates any subset of the components, at observation intervals longer than the time step of the ESN, with one ESN, and its ensemble spread estimates the uncertainty.

\subsection{Synchronization}\label{sec:sync_method}

In ensemble Kalman filters~\citep{evensen_sequential_1994,houtekamer2016review}, the ensemble statistics approximate the covariances of the Kalman update~\citep{kalman_new_1960}. We employ the EnSRKF, which updates the ensemble without perturbing the observations, and therefore avoids the sampling error of the stochastic filter with small ensembles~\citep{sakov_implications_2008}. {The filter updates the input and the reservoir state of every member together, i.e., the augmented state}
\begin{equation}
    \vect{\psi}_j(t_n) = \left[\vect{u}_j(t_n);\ \vect{r}_j(t_n)\right],
    \label{eq:augmented_state}
\end{equation}
in which the input is initially the readout of the reservoir state, Eq.~\eqref{eq:esn_update}. 
At an observation time, consider the ensemble $\matr{\Psi} = [\vect{\psi}_1,\ldots,\vect{\psi}_m]$, with mean $\bar{\vect{\psi}}$ and deviations
\begin{equation}
    \matr{\Psi}' = \left[\vect{\psi}_1-\bar{\vect{\psi}},\,\ldots,\,
          \vect{\psi}_m-\bar{\vect{\psi}}\right] = \left[\matr{U}';\,\matr{R}'\right],
    \label{eq:ensemble_anomaly}
\end{equation}
in which $\matr{U}'$ and $\matr{R}'$ are the deviations of the input and of the reservoir state, and we drop the time argument for brevity. The measurement operator, $\matr{M} \in \{0, 1\}^{N_q \times (N_u + N_r)}$, selects the measured components of the augmented state, e.g., $\matr{M} = [\mathbb{I}, \matr{0}]$ when every component is measured. Because the measured components are rows of the augmented state, $\matr{M}$ is linear, and the analysis requires no Jacobian of the ESN. The predicted observables have deviations $\matr{Q}' = \matr{M}\matr{\Psi}'$ and mean $\bar{\vect{q}} = \matr{M}\bar{\vect{\psi}}$. With the superscripts f and a for the forecast and the analysis, the analysis mean and deviations are
\begin{subequations}\label{eq:ensrkf}
\begin{equation}
    \bar{\vect{\psi}}^{\rm a} = \bar{\vect{\psi}}^{\rm f}
      + \matr{K}\underbrace{\big(\vect{d} - \bar{\vect{q}}^{\rm f}\big)}_{\text{innovation, }\vect{\nu}},
    \qquad
    \matr{\Psi}'^{\rm a} = \matr{\Psi}'^{\rm f}\,\matr{T}^{1/2},
    \label{eq:ensrkf_update}
\end{equation}
where the transform matrix, $\matr{T}$, and the Kalman gain, $\matr{K}$, are
\begin{equation}
  \begin{aligned}
    \matr{T} &= \Big[\mathbb{I}_m + \frac{1}{m-1}\,\big(\matr{Q}'^{\rm f}\big)\T\matr{C}_{dd}^{-1}\matr{Q}'^{\rm f}\Big]^{-1},\\
    \matr{K} &= \frac{1}{m-1}\,\matr{\Psi}'^{\rm f}\,\matr{T}\big(\matr{Q}'^{\rm f}\big)\T\matr{C}_{dd}^{-1},
  \end{aligned}
    \label{eq:gain}
\end{equation}
{in which} $\matr{C}_{dd}$ is the assumed observation-error covariance (\S\ref{sec:uq_method}), and $\matr{T}^{1/2}$ is the symmetric square root of $\matr{T}$~\citep{sakov_implications_2008}. 
{The forecast covariance of the predicted observables is}
\begin{equation}
    \matr{C}^{\rm f}_{qq} = \frac{1}{m-1}\,\matr{Q}'^{\rm f}\big(\matr{Q}'^{\rm f}\big)\T.
    \label{eq:covariances}
\end{equation}
\end{subequations}

The analysis synchronizes the ESN by correcting the reservoir state, which carries the memory of the ESN, {so that no washout is needed}. Under partial observations, the analysis also reconstructs the unmeasured inputs through their ensemble correlation with the measured inputs. As a baseline, which isolates the effect of correcting the reservoir state, we also deploy a filter that corrects only the input, i.e., the analysis updates $\vect{u}_j$ and leaves $\vect{r}_j$ at its forecast. The analysed input is then no longer the readout of the reservoir, and it corrects the reservoir only through $\matr{W}_\mathrm{in}$ at the next step (\S\ref{sec:sync}).

In the forecast, the input of every member is the readout of its reservoir state, {$\vect{u}_j = \matr{W}_\mathrm{out}[\vect{r}_j; 1]$} (Eq.~\eqref{eq:esn_update}). The analysis of Eq.~\eqref{eq:ensrkf_update} preserves this relation, because every analysis member is a combination of the forecast members with weights that sum to one, and the readout is affine in $\vect{r}$. Therefore, $\vect{u}^{\rm a}_j = \matr{W}_\mathrm{out}[\vect{r}^{\rm a}_j; 1]$, which we verify to round-off on every ESN of this paper with the global filter. The localized filter of \S\ref{sec:localization_method} breaks this relation.

The deployment costs $m$ closed-loop forecasts of the reservoir per time step and one analysis per observation. The cost of the analysis grows linearly with $N_r$, because the square root acts on the $m \times m$ matrix $\matr{T}$, and not on the covariance of the augmented state.

\subsection{Uncertainty quantification}\label{sec:uq_method}

The ensemble covariance, which the gain needs, also quantifies the uncertainty of the state and of the forecast, which a washout does not provide. The uncertainty is reliable only if the ensemble represents every source of error. Because every member is forecast by the same ESN and compared with the same data, the ensemble does not represent the noise in the data or the model error of the ESN.

The assumed observation-error covariance, $\matr{C}_{dd}$, accounts for the noise in the data, $\matr{C}_{\epsilon\epsilon}$, and for the model error of the ESN seen through the measurement operator, as the representation error does in numerical weather prediction~\citep{janjic_representation_2018}. With dense data, a $\matr{C}_{dd}$ that describes the noise alone is the most accurate. With sparse data, however, the analysis then attributes the whole innovation to the forecast, and the ensemble collapses (\S\ref{sec:uq_results}). We therefore set its standard deviation, $\sigma_\mathrm{obs}$, to four to ten times the noise in the data (Table~\ref{tab:cases}), which trades accuracy with dense data for robustness with sparse data.

Inflation enlarges the forecast deviations to counteract the loss of spread caused by sampling error and model error~\citep[Sect.~15.2]{evensen_data_2009}. Instead of tuning an inflation factor for each experiment, we estimate it from the innovation at every analysis~\citep{li_simultaneous_2009}. If the ensemble underestimates the forecast-error covariance by a factor $\gamma$, the expected squared innovation~\citep{desroziers_diagnosis_2005} estimates $\tilde{\gamma}$ from one analysis
\begin{equation}
    \mathbb{E}\big[\|\vect{\nu}\|^2\big] = \gamma\,\mathrm{tr}\,\matr{C}^{\rm f}_{qq} + \mathrm{tr}\,\matr{C}_{dd}
    \,\Rightarrow\,
    \tilde{\gamma} = \frac{\|\vect{\nu}\|^2 - \mathrm{tr}\,\matr{C}_{dd}}{\mathrm{tr}\,\matr{C}^{\rm f}_{qq}}.
    \label{eq:inflation}
\end{equation}
Therefore, the inflation grows only when the innovations exceed what $\matr{C}_{dd}$ and the spread explain, i.e., with the model error of the ESN, which the members do not carry. Because one innovation is a single sample, we bound $\tilde{\gamma}$ to $[1, 9]$, i.e., the inflation enlarges the deviations by at most a factor of three, and average it over approximately the last ten analyses, $\gamma \leftarrow 0.9\,\gamma + 0.1\,\tilde{\gamma}$, starting from $\gamma = 1$. While $\mathrm{tr}\,\matr{C}^{\rm f}_{qq}$ exceeds $\mathrm{tr}\,\matr{C}_{dd}$, i.e., before the ensemble synchronizes, we set $\tilde{\gamma} = 1$, because Eq.~\eqref{eq:inflation} assumes forecast errors that are small compared with the size of the attractor. The forecast ensemble is then inflated about its mean
\begin{equation}
    \vect{\psi}_j^{\rm f} \leftarrow \bar{\vect{\psi}}^{\rm f} + \sqrt{\gamma}\,\big(\vect{\psi}_j^{\rm f} - \bar{\vect{\psi}}^{\rm f}\big), \qquad j = 1, \ldots, m.
    \label{eq:inflate}
\end{equation}
Equation~\eqref{eq:inflate} applies one factor to every row of the augmented state except a physical parameter, also in the LETKF (\S\ref{sec:localization_method}). A physical parameter has no dynamics to restore its spread between analyses (\S\ref{sec:param_method}), so an inflated parameter drifts after every burst of model error (not shown). The filter that corrects only the input (\S\ref{sec:sync_method}) leaves the reservoir rows uninflated.

\subsection{Bimodal posterior}\label{sec:mixture_method}

The analysis of Eq.~\eqref{eq:ensrkf} approximates the forecast by one Gaussian. If the measured components are invariant under a symmetry of the system, the data do not distinguish the true state from its image under the symmetry. The posterior then has a mode at each, and one Gaussian analysis moves the members towards their mean, between the modes. For example, the Lorenz-63 system is invariant under the mirror symmetry $(x, y, z) \to (-x, -y, z)$, which leaves $z$ unchanged (\S\ref{sec:partial_results}). For such data, we employ a Gaussian-mixture ensemble filter~\citep{alspach_nonlinear_1972,bengtsson_toward_2003}, which fits up to two Gaussians to the forecast ensemble, updates the members of each Gaussian with their own mean and covariance, and reweights the Gaussians with the likelihood of the data. With one Gaussian, the filter is the EnSRKF of Eq.~\eqref{eq:ensrkf}. Details on the Gaussian-mixture ensemble filter are in Appendix~\ref{app:mixture}. 

\subsection{Parameter estimation}\label{sec:param_method}

The state can be augmented further with uncertain physical parameters of the system, which the filter then estimates together with the state. To this end, the ESN is trained at several values of a parameter, $\mu$. Parametric ESNs usually read the parameter as an additional input~\citep{kong_machine_2021}. Here, the ESN reads the parameter in the readout only
\begin{equation}
    \vect{u}(t_{n+1}) = \matr{W}_\mathrm{out}[\vect{r}(t_{n+1});\ 1] + \vect{w}^{\mu}_{\mathrm{out}}\,\tilde\mu,
    \label{eq:param_readout}
\end{equation}
where $\tilde\mu$ is the normalized parameter; and $\vect{w}^{\mu}_{\mathrm{out}} \in \mathbb{R}^{N_u}$ is its output weight, which is trained together with $\matr{W}_\mathrm{out}$. The reservoir does not read the parameter, which reaches it only through the closed loop. This choice has two consequences (Appendix~\ref{app:readout_param}). First, the washout needs no value of the parameter, so the members start as in state estimation. Second, a member whose parameter does not match its reservoir state stays on the attractor, also with weak regularization. The ESN is deployed with $\mu_j$ appended to the augmented state of each member (Fig.~\ref{fig:schematic}b, navy)
\begin{equation}
    \vect{\psi}_j = \left[\vect{u}_j;\ \vect{r}_j;\ \mu_j\right].
    \label{eq:augmented_param}
\end{equation}
The parameter has no dynamics: $\mu_j$ is constant in the forecast and changes only at the analyses, through its ensemble correlation with the observables. In \S\ref{sec:param_results}, the parameter is the forcing, $F$, of the Lorenz-96 system.

\subsection{High-dimensional systems}\label{sec:localization_method}

\begin{figure}
  \centering
  \includegraphics[width=\columnwidth]{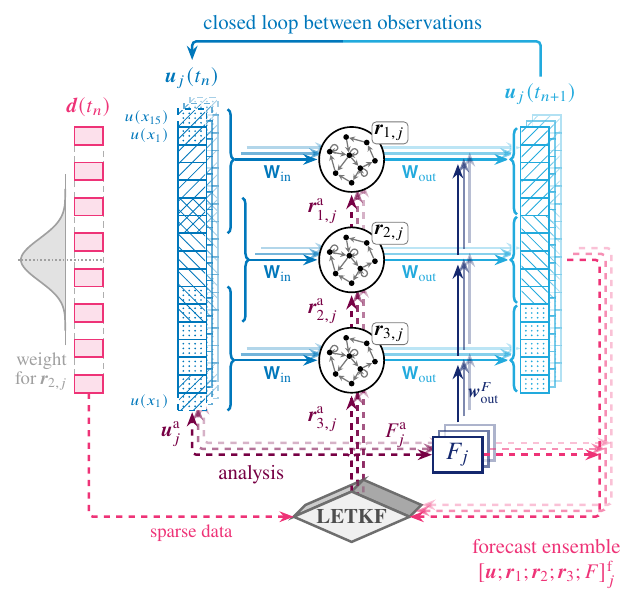}
  \caption{Data assimilation with a parallel ESN of three reservoirs and the local ensemble transform Kalman filter (LETKF), on a ring of $15$ sites. Each reservoir reads the five sites of its patch and a halo of one site on either side (hatching), and forecasts the sites of its patch; the reservoirs share their matrices. Left: the data at every other site (filled), and the Gaspari--Cohn weight (light grey) with which the LETKF assimilates them in the analysis of $\vect{r}_{2,j}$. Navy: the parameter of each member of a parametric ESN, $F_j$, which the readout of every reservoir reads through the shared $\vect{w}^{F}_{\mathrm{out}}$ (\S\ref{sec:param_method}). Other elements as in Fig.~\ref{fig:schematic}b.}
  \label{fig:parallel_schematic}
\end{figure}

A single reservoir reads and forecasts every site of the system, so its size, and the size of the augmented state, Eq.~\eqref{eq:augmented_state}, grow with the size of the system. A parallel ESN~\citep{pathak_model_2018,vlachas_backpropagation_2020} splits the domain into patches. The reservoir of each patch reads the sites of its patch and a halo of neighbouring sites, and forecasts the sites of its patch (Fig.~\ref{fig:parallel_schematic}). When the dynamics are invariant under a shift of the sites, as in the Lorenz-96 system, the reservoirs share $\matr{W}_\mathrm{in}$, $\matr{W}$ and $\matr{W}_\mathrm{out}$, and $\matr{W}_\mathrm{out}$ is trained on all the patches at once. The augmented state stacks the input and the reservoir states of the $P$ patches
\begin{equation}
    \vect{\psi}_j(t_n) = \left[\vect{u}_j(t_n);\ \vect{r}_{1,j}(t_n);\ \ldots;\ \vect{r}_{P,j}(t_n)\right],
    \label{eq:augmented_parallel}
\end{equation}
which has $N_u + P N_r$ rows, e.g., $4040$ rows for the parallel ESN of the 40-site Lorenz-96 system (Table~\ref{tab:esns}), against $m = 5$ to $160$ members.

{An ensemble filter can be applied to a high-dimensional system by first encoding the state onto a low-dimensional space, in which the ESN forecasts and the filter updates the reduced state~\citep{ozalp_real-time_2026}. In the full space,} localization (\S\ref{sec:intro}) tapers the covariance between a row of the state and an observation with the Gaspari--Cohn function of their distance~\citep{gaspari1999}. We employ the local ensemble transform Kalman filter (LETKF)~\citep{hunt2007efficient}, in which every row, $i$, of the augmented state is updated with its own transform. The analysis mean and deviations of row $i$ are
\begin{subequations}\label{eq:letkf_all}
\begin{equation}
    \bar{\psi}_i^{\rm a} = \bar{\psi}_i^{\rm f} + \vect{k}_i\,\vect{\nu},
    \qquad
    \vect{\psi}_i'^{\rm a} = \vect{\psi}_i'^{\rm f}\,\matr{T}_i^{1/2},
    \label{eq:letkf}
\end{equation}
where $\vect{\psi}_i'$ is the $i$-th row of $\matr{\Psi}'$ , and {the transform matrix, $\matr{T}_i$, and the Kalman gain, $\vect{k}_i$, of row $i$ are those of Eq.~\eqref{eq:gain} with the tapered precision, $\matr{L}_i\,\matr{C}_{dd}^{-1}$, in place of $\matr{C}_{dd}^{-1}$}
\begin{equation}
  \begin{aligned}
    \matr{T}_i &= \Big[\mathbb{I}_m + \frac{1}{m-1}\,\big(\matr{Q}'^{\rm f}\big)\T\matr{L}_i\,\matr{C}_{dd}^{-1}\matr{Q}'^{\rm f}\Big]^{-1},\\
    \vect{k}_i &= \frac{1}{m-1}\,\vect{\psi}_i'^{\rm f}\,\matr{T}_i\big(\matr{Q}'^{\rm f}\big)\T\matr{L}_i\,\matr{C}_{dd}^{-1},
  \end{aligned}
    \label{eq:letkf_gain}
\end{equation}
where $\matr{T}_i^{1/2}$ is the symmetric square root of $\matr{T}_i$, and 
\begin{equation}
    \matr{L}_i = \mathrm{diag}\big(\ell(\delta_{i1}/h), \ldots, \ell(\delta_{iN_q}/h)\big)  
    \label{eq:letkf_taper}
\end{equation}
\end{subequations}
is the taper that weights the precision of observation $k$ with the Gaspari--Cohn function, $\ell$, which vanishes for $\delta_{ik} \geq 2h$; $\delta_{ik}$ is the distance between the position of row $i$ and observation $k$; and $h$ is the half-width. Because $\matr{C}_{dd}$ is diagonal, the product $\matr{L}_i\matr{C}_{dd}^{-1}$ is symmetric. Without localization, $\matr{L}_i = \mathbb{I}$, and Eq.~\eqref{eq:letkf_all} reduces to Eq.~\eqref{eq:ensrkf} row by row: $\matr{T}_i = \matr{T}$, and $\vect{k}_i$ is row $i$ of $\matr{K}$; we call this case the global filter. Rows at the same position share $\matr{L}_i$, and hence $\matr{T}_i$, so the analysis requires one $m \times m$ transform per position.

Localization requires a position for every row of the augmented state. An input row is at its site, and the reservoir state of patch $p$ of a parallel ESN, $\vect{r}_{p,j}$, at the centre of its patch (Fig.~\ref{fig:parallel_schematic}), as in \citet{penny_integrating_2022}. The input rows of a patch and its reservoir state are at different positions, and are therefore updated with different transforms. Hence, the analysed input is no longer the readout of the analysed reservoir state, $\vect{u}^{\rm a}_{p,j} \neq \matr{W}_\mathrm{out}[\vect{r}^{\rm a}_{p,j}; 1]$, where $\vect{u}_{p,j}$ collects the sites of patch $p$. We keep both: at the next step, the reservoir reads the analysed input, which is closer to the data than the readout (\S\ref{sec:loc_results}).

\section{Test cases and implementation}
\label{sec:experiments}

Table~\ref{tab:esns} lists the six ESNs, which are trained on the Lorenz-63~\citep{lorenz_deterministic_1963} and Lorenz-96~\citep{lorenz_predictability_1996} systems, with their training settings. Table~\ref{tab:cases} lists the data and the data assimilation of each system. The Lorenz-63 system, which is a model of thermal convection reduced to three modes, reads
\begin{equation}
    \frac{\mathrm{d}x}{\mathrm{d}t} = \sigma\,(y - x), \quad
    \frac{\mathrm{d}y}{\mathrm{d}t} = x\,(\rho - z) - y, \quad
    \frac{\mathrm{d}z}{\mathrm{d}t} = x\,y - b\,z,
    \label{eq:lorenz63}
\end{equation}
where $x$, $y$ and $z$ are the state; and $\sigma = 10$, $\rho = 28$ and $b = 8/3$ are the Prandtl number, the normalized Rayleigh number and a geometric factor, for which the system is chaotic. The Lorenz-96 system is a ring of $N_u$ sites
\begin{equation}
    \frac{\mathrm{d}u_i}{\mathrm{d}t} = \left(u_{i+1} - u_{i-2}\right) u_{i-1} - u_i + F, \qquad i = 1, \dots, N_u,
    \label{eq:lorenz96}
\end{equation}
where $u_i$ is the state at site $i$, with periodic indices, $u_{i \pm N_u} = u_i$. The quadratic term transports the state along the ring, the linear term, $-u_i$, dissipates it, and the forcing, $F$, drives it. With $N_u = 10$ and $40$ sites, the system is chaotic at $F = 8$.

\begin{table}
  \centering
  \footnotesize
  \setlength{\tabcolsep}{4pt}
  \caption{Data and data assimilation of the test cases. Noise: standard deviation of the noise in the data, relative to that of each component; $\sigma_\mathrm{obs}$: standard deviation of $\matr{C}_{dd}$, relative to that of each observable. Steps are ESN steps, of size $\Delta t$, and $T_\lambda$ is that of the system at $F = 8$ for Lorenz-96; the observation interval is given in steps and in $T_\lambda$. The results vary the observation interval (\S\ref{sec:results}).}
  \label{tab:cases}
  \begin{tabular}{@{}lccc@{}}
    \toprule
     & & \multicolumn{2}{c}{Lorenz-96} \\
    \cmidrule(lr){3-4} 
    Description & Lorenz-63 & $N_u = 10$ & $N_u = 40$ \\ \midrule
    Time step of the record, $\delta t$ & $0.02$ & $0.01$ & $0.01$ \\
    Noise in the data & $5\%$ & $2\%$ & $2\%$ \\
    ESN time step, $\Delta t$; steps per $T_\lambda$ & $0.04$; $27$ & $0.02$; $42$ & $0.02$; $30$ \\
    Upsampling, $\Delta t / \delta t$ & \multicolumn{3}{c}{$2$} \\
    Observation interval, $\Delta t_\mathrm{obs}$ & $5$; $0.18$ & $4$; $0.095$ & $5$; $0.17$ \\
    Assumed observation error, $\sigma_\mathrm{obs}$ & \multicolumn{3}{c}{$0.2$} \\
    \bottomrule
  \end{tabular}
\end{table}

Time is measured in Lyapunov times, $T_\lambda = 1/\lambda_1$, where $\lambda_1$ is the leading Lyapunov exponent. Each experiment starts $2\,T_\lambda$ into a test record, which is not seen in training, and is repeated over five windows shifted by $4\,T_\lambda$ ($20$ windows in \S\ref{sec:sync}, ten for the parallel ESNs, and $38$ for L96-40 in \S\ref{sec:loc_results}). The comparison of the deployments of L63 is repeated over $100$ initializations (\S\ref{sec:partial_results}), and the parameter estimation over seven forcings and nine observation intervals, i.e., $63$ runs (\S\ref{sec:param_results}). Every record carries Gaussian noise, whose standard deviation is that of Table~\ref{tab:cases} relative to that of each component. The ESNs are, therefore, trained on noisy records, and the washout and the filter read noisy data. The noise-free record serves only to score the deployments (\S\ref{sec:results}). The assumed observation-error covariance, $\matr{C}_{dd}$, is diagonal.

The hyperparameters of the ESNs are selected with the recycle validation of \citet{racca_robust_2021}, with a validation error that weights the short-term prediction, i.e., the accuracy of the closed-loop forecast, and the long-term prediction, i.e., the stability and the statistics of a long closed loop~\citep{platt_systematic_2022}. Each unit reads one input component and a constant input equal to the mean magnitude of the normalized input~\citep{racca_robust_2021}, which gives the input bias, $\vect{b}_\mathrm{in}$. The halo of the parallel ESNs needs at least two sites, because the tendency of site $i$ depends on sites $i-2$ to $i+1$ (Eq.~\eqref{eq:lorenz96}). Appendix~\ref{app:verification} lists the selected hyperparameters and the ranges of the search, and verifies the ESNs.

\begin{algorithm}
  \caption{Washout-free deployment of a trained ESN.}
  \label{alg:deployment}
  \begin{algorithmic}[1]
    \Require trained ESN, Eqs.~\eqref{eq:esn}--\eqref{eq:ridge}; data $\vect{d}(t_n)$ from the first observation, $t = 0$, to the last, $t_\mathrm{end}$; $\matr{C}_{dd}$; $m$
    \State Draw $m$ members, $\vect{\psi}_j$, from validation; $\gamma \gets 1$
    \For{every observation time $t_n \le t_\mathrm{end}$}
      \State {$\vect{\psi}^{\rm f}_j \gets \textsc{Forecast}(\vect{\psi}^{\rm a}_j,\, t_n)$, $j = 1, \ldots, m$}: closed loop, Eqs.~\eqref{eq:reservoir}--\eqref{eq:esn_update}
      \State {$\gamma \gets \textsc{Inflation}\big(\gamma,\, \vect{d}(t_n) - \matr{M}\vect{\psi}^{\rm f}_j,\, \matr{C}_{dd}\big)$, Eq.~\eqref{eq:inflation}; \State  $\vect{\psi}^{\rm f}_j \gets \textsc{Inflate}(\vect{\psi}^{\rm f}_j,\, \gamma)$, Eq.~\eqref{eq:inflate}}
      \State {$\vect{\psi}^{\rm a}_j \gets \textsc{Analysis}(\vect{\psi}^{\rm f}_j,\, \vect{d}(t_n),\, \matr{C}_{dd})$}  Eq.~\eqref{eq:ensrkf} or~\eqref{eq:letkf_all}
    \EndFor
    \State {$\vect{\psi}_j \gets \textsc{Forecast}(\vect{\psi}^{\rm a}_j,\, t > t_\mathrm{end})$, $j = 1, \ldots, m$}
  \end{algorithmic}
\end{algorithm}

Algorithm~\ref{alg:deployment} summarizes the deployment. Each member starts from a reservoir state that the ESN visits in open loop on a random window of the validation data, i.e., the part of the training record held out for the selection of the hyperparameters. The members, therefore, start on the learned attractor, with the readout as their input. These states are computed once from the training record. The reservoir of the parametric ESN does not read the parameter (\S\ref{sec:param_method}), so its members start in the same way, from windows of the records of every training forcing. Each member draws its parameter uniformly over the trained range, independently of its window. The analysis is not constrained to the learned attractor. Where the data do not determine the state, the analysis can move members off the attractor, from which the closed loop diverges (\S\ref{sec:partial_results}).

\section{Results}
\label{sec:results}

\begin{figure*}
  \centering
  \includegraphics[width=\textwidth]{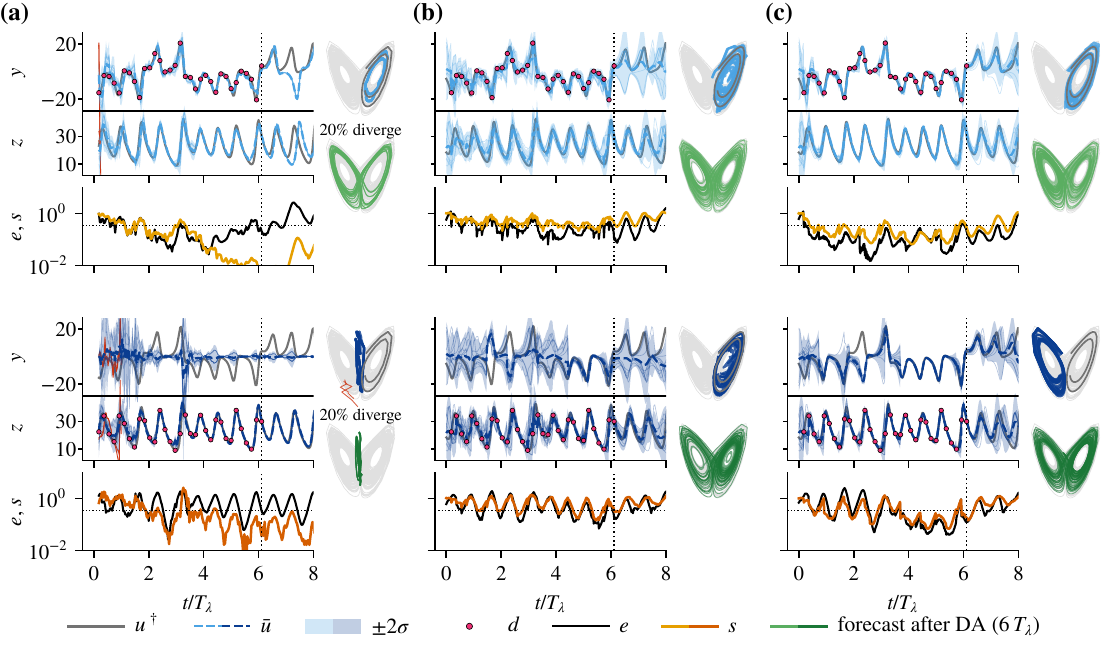}
  \caption{{Deployment of L63} from one observed component (top: $y$; bottom: $z$), on one test window.
  (a) Direct replacement in the partial-input ESN, and the EnSRKF with  $m = 10$ members that corrects (b) the input and (c) the input and the reservoir with the Gaussian-mixture analysis for the $z$-only case. 
  Data is observed every $0.18\,T_\lambda$ for $6\,T_\lambda$ (dotted line), then the ESNs are run in closed loop. 
  Left:  $y$ and $z$ in time, with the members, the ensemble mean, $\bar{u}$, its $\pm2\sigma$ band, the true state, $u^\dagger$ (grey), and the observations, $d$; below, the error, $e$ (black), and the spread, $s$ (oranges), medians over $20$ ensembles; dotted: the threshold of $\tau_\mathrm{e}$ and $\tau_\mathrm{s}$ (Eq.~\eqref{eq:sync_time}). 
  Right: the members in the $x$--$z$ plane over the last $1.3\,T_\lambda$ of the data (blue) and over the closed loop (green), on the training attractor (light grey); red: diverging members, excluded from the statistics.}
  \label{fig:l63_forecast}
\end{figure*}

We score every deployment against the true state, $\vect{u}^\dagger$, with the error, $e$, of the ensemble mean and the spread, $s$, of the ensemble
\begin{gather}
  e(t_n) = \frac{\|\bar{\vect{u}}(t_n)-\vect{u}^\dagger(t_n)\|}{\sigma^\dagger}, \;
  s(t_n) = \frac{\sqrt{\mathrm{tr}\,\matr{C}_{uu}(t_n)}}{\sigma^\dagger}, 
  \label{eq:error_spread}
\end{gather}
where $\sigma^\dagger = \big\langle\|\vect{u}^\dagger-\langle\vect{u}^\dagger\rangle\|^2\big\rangle^{1/2}$ is the standard deviation of the true state; 
$\bar{\vect{u}}$ is the ensemble mean of the input; $\matr{C}_{uu} = \matr{U}'\big(\matr{U}'\big)\T/(m-1)$ is its ensemble covariance, with the deviations $\matr{U}'$ of Eq.~\eqref{eq:ensemble_anomaly}; and $\langle\cdot\rangle$ is the time average over the test record. The error and the spread share one normalization, so that an error of one corresponds to forecasting the time-averaged state, i.e., climatology. The times at which the error and the spread first fall below $0.35$, about a third of the error of climatology, are
\begin{gather}
  \tau_\mathrm{e} = \min_{t_n\le t_\mathrm{end}}\big\{t_n : e(t_n)<0.35\big\}, \\
  \tau_\mathrm{s} = \min_{t_n\le t_\mathrm{end}}\big\{t_n : s(t_n)<0.35\big\},
  \label{eq:sync_time}
\end{gather}
where time is measured from the first observation, and $t_\mathrm{end}$ is the time of the last observation. The synchronization time, $\tau_\mathrm{s}$, needs no true state, whereas $\tau_\mathrm{e}$ is its reference. The \emph{final error}, $e_\mathrm{end}$, scores the synchronized estimate, and the \emph{valid time}, $\tau_\mathrm{v}$, the forecast after the data
\begin{gather}
  e_\mathrm{end} = \big\langle e^2 \big\rangle^{1/2}_{(t_\mathrm{end}-T_\lambda,\; t_\mathrm{end}]}, \\
  \tau_\mathrm{v} = \min_{t_n>t_\mathrm{end}}\big\{t_n-t_\mathrm{end} : e(t_n)>0.5\big\},
  \label{eq:success}
\end{gather}
where the subscript of $\langle\cdot\rangle$ is the interval of the average, i.e.,  the last Lyapunov time of the data, or all the data for records shorter than $T_\lambda$. A deployment fails when $e_\mathrm{end} > 0.5$. This threshold, which also defines the valid time, is half the error of climatology. The \emph{spread--error ratio}, $\langle s^2\rangle^{1/2}/\langle e^2\rangle^{1/2}$ from $2\,T_\lambda$ after the first observation to $t_\mathrm{end}$, i.e.,  after the synchronization transient, is close to one for a calibrated ensemble, below one for an under-dispersed ensemble, and above one for an over-dispersed ensemble.

\subsection{Synchronization}
\label{sec:deployment}\label{sec:sync}\label{sec:partial_results}

Figure~\ref{fig:l63_forecast} (top) compares three deployments of L63 from $y$ alone: (a) direct replacement in the partial-input ESN (\S\ref{sec:direct_method}); (b) the filter that corrects only the input; and (c) the filter that corrects the input and the reservoir (\S\ref{sec:sync_method}). First, the realizations of direct replacement converge onto one trajectory by the echo state property, so their spread, an eighth of their error, carries no information on the error. Second, correcting only the input synchronizes the ESN in part, with twice the error of the filter that also corrects the reservoir, because the reservoir stays where the forecast left it. Third, the filter that corrects the input and the reservoir synchronizes the ESN within $0.4\,T_\lambda$, with the smallest error of the three and a spread that exceeds its error.

\begin{figure}[pos=!htbp]
  \centering
  \includegraphics[width=\columnwidth]{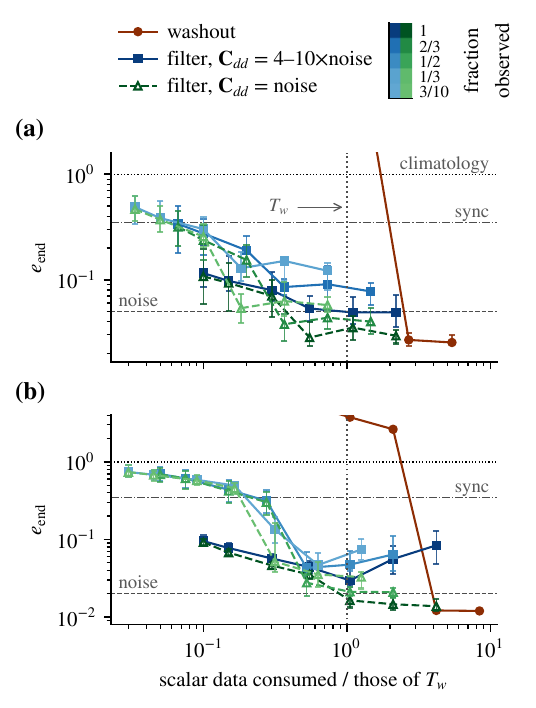}
  \caption{Data required for synchronization for (a) L63 and (b) L96-10s. Final error, $e_\mathrm{end}$ (Eq.~\eqref{eq:success}), against the number of scalar data, relative to that of the training washout of the full state, $T_w N_u$, medians and interquartile ranges (error bars) over $20$ windows; sync: $e = 0.35$, the threshold of synchronization (Eq.~\eqref{eq:sync_time}). Dark orange: washout of the full state at every ESN step, from $\vect{r} = \vect{0}$. Blues: the filter with $\matr{C}_{dd}$ of Table~\ref{tab:cases}. Greens: the filter with $\matr{C}_{dd}$ equal to the noise in the data. The filter has $m = 30$ on L63 and $64$ on L96-10s.}
  \label{fig:sync}
\end{figure}

Figure~\ref{fig:sync} measures the data that the washout and the filter consume before the ESN forecasts, over $20$ windows.  
\begin{figure}[pos=!t]
  \centering
  \includegraphics[width=\columnwidth]{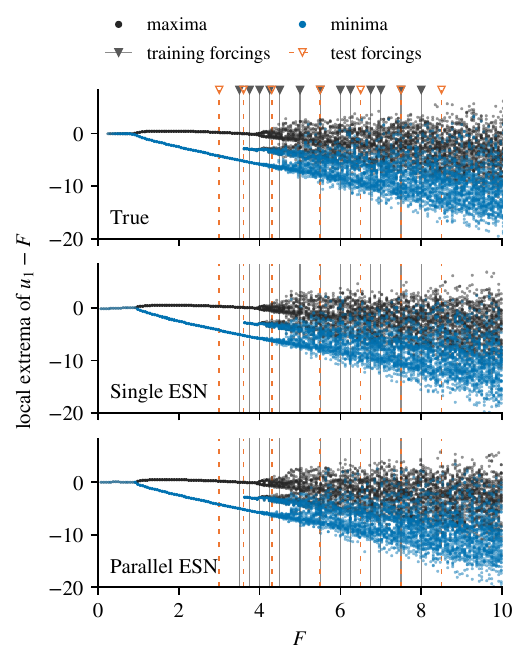}
  \caption{Bifurcation diagrams of the ten-site Lorenz-96 system (top, True), of the parametric L96-10s (middle, Single ESN) and of a parallel parametric ESN (bottom, Parallel ESN) in closed loop: local extrema in time of the first site relative to the fixed point, $u_1 - F$. Each ESN runs at the forcings of the true system, over the last $35$ of $400$ time units, from one washout per forcing on a random window of the validation data. Grey lines with filled triangles: training forcings of the parametric L96-10s; dashed lines with open triangles: test forcings.}
  \label{fig:bifurcation}
\end{figure}
With the full state observed, three analyses synchronize both ESNs, i.e., $9$ data on L63 and $30$ data on L96-10s, which are $15\%$ of the data of the training washout, whereas a washout of as many data does not. The forecast after these analyses is valid for $0.9\,T_\lambda$ on L63 and $1.8\,T_\lambda$ on L96-10s. A washout of a quarter of a Lyapunov time, however, already forecasts as long as the filter or longer. The gain of the filter is, therefore, with few data and with partial observations, from which a washout cannot run. With $2\,T_\lambda$ of data or more, the filter synchronizes the ESN in every window from one and two components of L63, and from five and three of the ten sites of L96-10s. With more data, the final error of the filter is $0.05$ on L63 and $0.03$ to $0.08$ on L96-10s, above that of the washout, $0.03$ and $0.012$, because the analysis weighs the data with $\matr{C}_{dd}$. With $\matr{C}_{dd}$ equal to the noise in the data, the error of the filter is $0.03$ on L63 and $0.015$ on L96-10s (Fig.~\ref{fig:sync}, greens).

We repeat the comparison of Fig.~\ref{fig:l63_forecast} from every subset of the components of L63 over $100$ initializations, with $m = 5$ to $30$ members. The filter synchronizes the ESN from every subset except $z$ alone in at least $96\%$ of the runs (at least $29\%$ when updating the input only), whereas direct replacement fails in up to $50\%$ of the runs. From $z$ alone, the sign of $x$ and $y$ is not observable (\S\ref{sec:mixture_method}), so we score the runs up to the symmetry, i.e., we mirror every member that is closer to the mirror image of the true state than to the true state. With one Gaussian, $52$ to $85\%$ of the members leave the attractor, and the filter synchronizes the ESN in only $16$ to $46\%$ of the runs. With the Gaussian-mixture analysis (Fig.~\ref{fig:l63_forecast}c, bottom), the filter synchronizes the ESN in $61$, $83$ and $91\%$ of the runs with $m=10$, $20$ and $30$, whereas direct replacement fails in $96\%$ of the runs.
Similarly, on L96-10s, with $20\,T_\lambda$ of data, direct replacement diverges in every window from five, three and two sites, because nothing corrects the unmeasured components. In contrast, the filter synchronizes the ESN in every window, with a median final error of $0.07$, $0.12$ and $0.11$ from five, three and two sites, respectively.

\subsection{Parameter estimation}
\label{sec:param_results}

\begin{figure}
  \centering
  \includegraphics[width=\columnwidth]{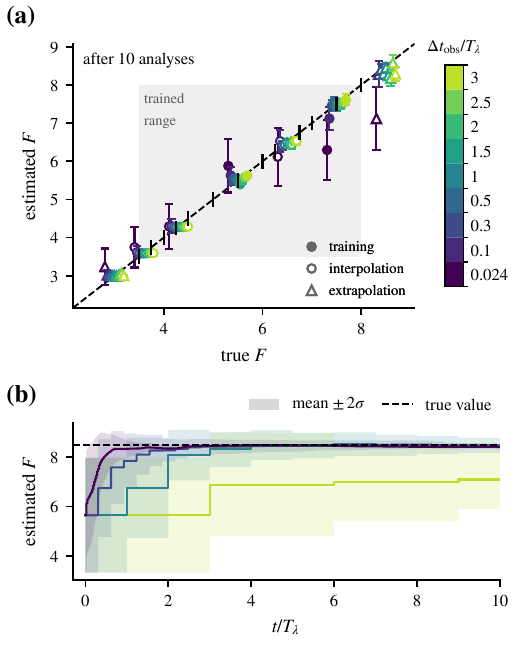}
  \caption{State and parameter estimation with the parametric L96-10s and the global filter, with $m = 64$. (a) Estimated against true forcing. (b) Estimated forcing against time for $F = 8.5$, beyond the trained range, with data every $\Delta t_\mathrm{obs} = 0.024$, $0.3$, $1$ and $3\,T_\lambda$ (colours as in (a)).}
  \label{fig:param}
\end{figure}

\begin{figure}
  \centering
  \includegraphics[width=\columnwidth]{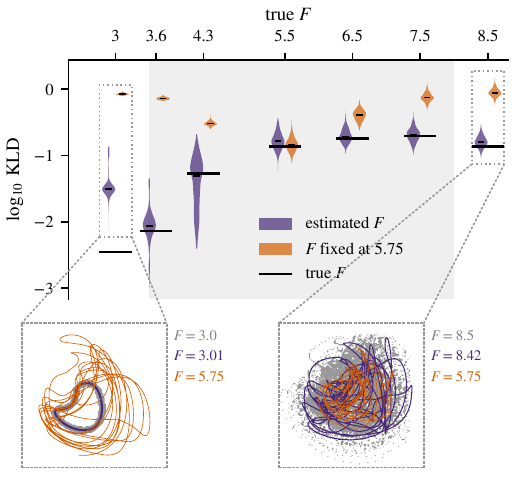}
  \caption{Forecast of the parametric L96-10s after the estimation of Fig.~\ref{fig:param} ($m = 64$): Kullback--Leibler divergence (KLD) of the closed-loop forecast over the $30\,T_\lambda$ after the last observation. $F$ fixed: state estimation only; true $F$: the true equations, from the noisy state at the end of the data. Insets: the extrapolation cases in the $u_1$--$u_2$ plane.}
  \label{fig:kld}
\end{figure}

We train the parametric L96-10s with $1000$ units and a parametric L96-10p with two reservoirs of $500$ units (Fig.~\ref{fig:parallel_schematic}, navy) at $13$ forcings, $F = 3.5$ to $8$, which cross a limit cycle, a quasi-periodic band and chaos (Fig.~\ref{fig:bifurcation}, top). In closed loop, both ESNs reproduce these regimes, also beyond the trained range (Fig.~\ref{fig:bifurcation}, middle and bottom). Their parameter estimation is comparable (not shown), so we report the single ESN. The global filter, with $m = 64$, estimates $[\vect{u}_j;\ \vect{r}_j;\ F_j]$ (Fig.~\ref{fig:schematic}b) from every site, for $30\,T_\lambda$, at seven forcings: two training forcings, three between the training forcings, and two beyond the trained range. The members start from states visited during training, with forcings drawn uniformly over the trained range (\S\ref{sec:experiments}). These states must cover the attractors of the trained range: from the limit cycle at $F = 3.5$ alone, the filter estimates the forcing within $2\%$ in $9$ of $30$ chaotic runs, against $29$ of $30$.

The forcing in the readout (\S\ref{sec:param_method}) compares favourably with the forcing in the input~\citep{kong_machine_2021} on ESNs of the same size (Appendix~\ref{app:readout_param}). Both ESNs estimate the forcing with comparable accuracy, but the output is about $40$ times less sensitive to the forcing in the readout. Over closed loops at forcings from $0.06$ to $10$, $1$ in $825$ leaves the attractor with the forcing in the readout, against $1$ in $7$ with the forcing in the input, and none against more than half with a Tikhonov factor of $10^{-12}$. Moreover, the washout needs no value of the forcing.

The filter recovers the forcing with a median error, at the last analysis, of $0.2\%$ inside the trained range, and of $0.2\%$ at $F = 3$ and $1.2\%$ at $F = 8.5$, to which the ESN extrapolates (Fig.~\ref{fig:param}a). With data every $0.1$ to $1\,T_\lambda$, the error is below $2\%$ at every forcing. Sparser data do not constrain the forcing in chaos, i.e., for $F \geq 5.5$: the error reaches $3.5\%$ at $F = 8.5$ observed every $2\,T_\lambda$. With data at every step, the spread of the forcing collapses (Fig.~\ref{fig:param}b), because the forcing is neither inflated nor driven by dynamics, so every analysis shrinks its spread. With $\matr{C}_{dd}$ equal to the noise, the $2\sigma$ band covers the true forcing in $9$ of the $63$ runs, against $38$ with the $\matr{C}_{dd}$ of Table~\ref{tab:cases}.

The estimated forcing lengthens the forecast of the ESN. The forecast is valid for the whole closed loop of $30\,T_\lambda$ on the limit cycle and the quasi-periodic band, and for up to $7.6\,T_\lambda$ in chaos with data every $1\,T_\lambda$ or more often, against at most $0.7\,T_\lambda$ with the forcing held at the centre of the trained range. The Kullback--Leibler divergence (KLD) between the probability densities of the true system and of each member is $2$ to $83$ times larger with the forcing held (Fig.~\ref{fig:kld}). In chaos, the KLD at the estimated forcing is within $0.08$ in $\log_{10}$ of that of the true equations started from the same data. At $F = 4.3$, in the quasi-periodic band, the KLD of the members spans almost two orders of magnitude, which is consistent with each member sampling a different part of the torus over the $30\,T_\lambda$ of the forecast.

\subsection{Localization}
\label{sec:loc_results}

\begin{figure}
  \centering
  \includegraphics[width=\columnwidth]{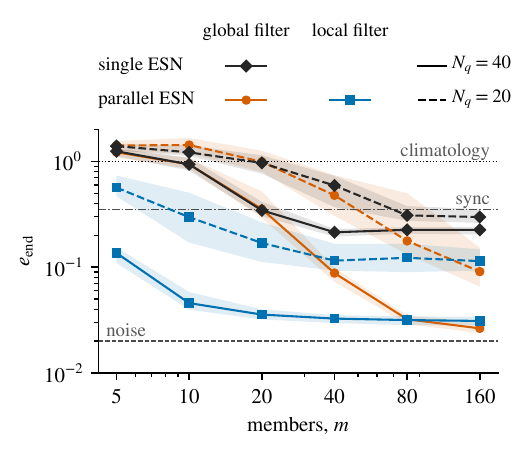}
  \caption{Localization on L96-40: final error, $e_\mathrm{end}$, against the number of members, $m$. The parallel ESN, L96-40p, is filtered by the global filter (vermilion) and by the LETKF with a half-width of two sites (blue), and the single ESN, L96-40s, by the global filter (black), from every site ($N_q = 40$, solid) and every other site ($N_q = 20$, dashed), observed every $0.17\,T_\lambda$ for $2\,T_\lambda$; sync: $e = 0.35$ (Eq.~\eqref{eq:sync_time}).}
  \label{fig:localization}
\end{figure}

Figure~\ref{fig:localization} shows how many members the deployment of L96-40p needs. The augmented state has $4040$ rows, whereas the ensemble spans at most $m - 1$ directions. With every site observed, the global filter fails in every window with $5$ and $10$ members, and synchronizes the ESN in every window from $40$ members, whereas the LETKF synchronizes the ESN in every window from $5$ members. From $20$ members on, more members do not lower the error of the LETKF, $0.031$ to $0.036$. With $160$ members, the global filter is slightly more accurate than the LETKF, with an error of $0.026$, which is consistent with the taper also removing real correlations beyond its half-width. With every other site observed, the LETKF fails in $3$ and $2$ of the $38$ windows with $20$ and $40$ members, whereas the global filter fails in $10$ and $6$ with $80$ and $160$ members. With $\sigma_\mathrm{obs}$ equal to the noise in the data, the error of the LETKF with every site observed falls to $0.015$, but $5$ members fail in every window with every other site observed, against $25$ of the $38$ windows with the $\sigma_\mathrm{obs}$ of Table~\ref{tab:cases}.

The global filter needs many members because a small ensemble samples spurious long-range correlations, which the taper of the LETKF removes (Appendix~\ref{app:covariance}).

\citet{penny_integrating_2022} paired a parallel ESN with the LETKF on the same 40-site system, with an ensemble initialized by a washout of $10$ to $20$ model time units of the noisy full state. Here, the members start from states visited during training, with no washout on the deployed system.

\subsection{Uncertainty quantification}
\label{sec:uq_results}

\begin{figure*}
  \centering
  \includegraphics[width=\textwidth]{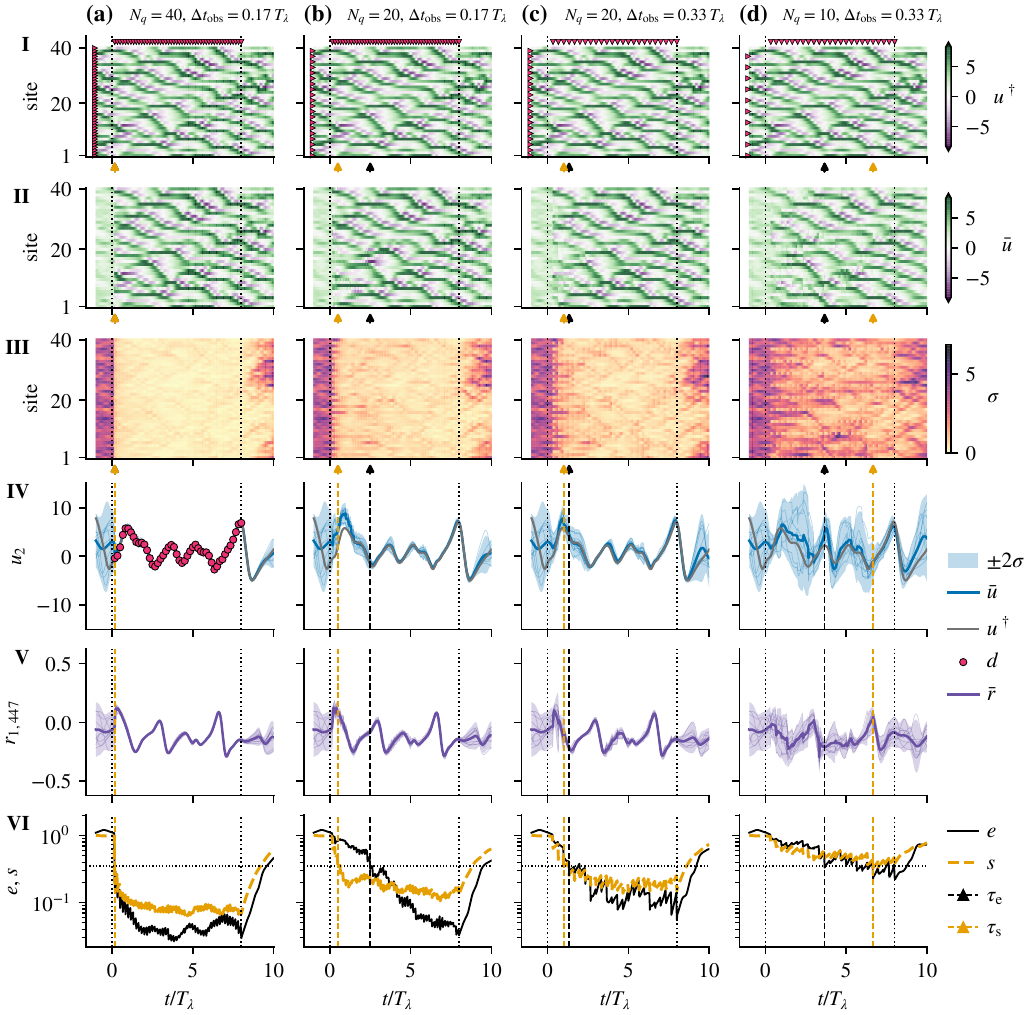}
  \caption{{Synchronization and uncertainty of L96-40p.} The LETKF, with $m = 10$ and a half-width of two sites, is deployed over the assimilation window (dotted lines), after $1\,T_\lambda$ of closed loop without data ($t < 0$), and the ESN forecasts in closed loop beyond it, in the window closest to the median valid times of ten windows. (a)--(d) Sensor layouts, with the observed sites, $N_q$, and the observation interval, $\Delta t_\mathrm{obs}$; $8\,T_\lambda$ of data; triangles: observed sites and analysis times. (I) True state, $u^\dagger$, (II) ensemble mean, $\bar{u}$, and (III) standard deviation, $\sigma$; (IV) site $2$, with the members, $\bar{u}$ and its $\pm2\sigma$ band, $u^\dagger$ and the observations, $d$; (V) one unit of reservoir $1$, which predicts sites $1$ to $5$, with the members, their mean, $\bar{r}$, and $\pm2\sigma$ band; (VI) $e$ and $s$, with the first times, $\tau_\mathrm{e}$ and $\tau_\mathrm{s}$, at which the error and the spread fall below $0.35$ (dotted): dashed lines, and arrows under I--III.}
  \label{fig:parallel_forecast}
\end{figure*}

The ensemble spread signals, without the true state, when the ESN synchronizes and when the filter fails. On L96-40p, Fig.~\ref{fig:parallel_forecast} shows the LETKF for four sensor layouts with decreasing data. With data, the members of a reservoir unit collapse onto one trajectory, i.e., the reservoir synchronizes with no washout (Fig.~\ref{fig:parallel_forecast}, V). The synchronization time, $\tau_\mathrm{s}$, grows as the data become sparser, from $0.17$ to $1.0\,T_\lambda$ (medians) in the three denser layouts. From every fourth site, the spread stays above $0.35$ for $5.5\,T_\lambda$ (median), and the filter fails in six of the ten windows, because ten sites constrain the 40-site ring only in part. On L63, a failed synchronization shows as a member that leaves the attractor, or as a spread above $0.5$ over the last Lyapunov time of the data. From $z$ alone, with the Gaussian-mixture analysis and $20$ to $30$ members, either signal occurs in $76$ to $89\%$ of the runs that fail. In about half of the runs that do not fail, the ensemble keeps members on both modes, and the spread then measures the distance between the modes.

\begin{figure}
  \centering
  \includegraphics[width=\columnwidth]{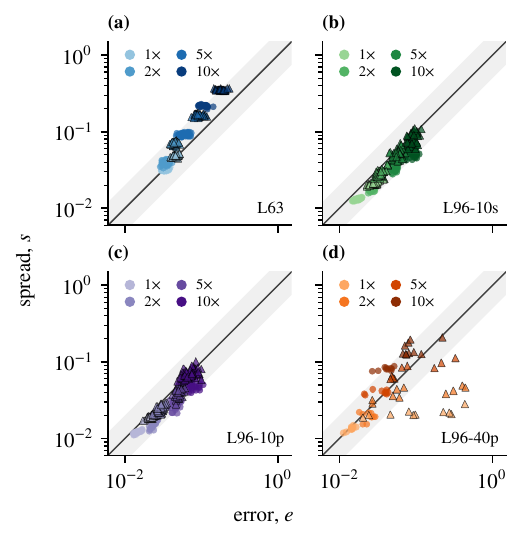}
  \caption{Ensemble spread against the error of the ensemble mean on (a) L63, (b) L96-10s, (c) L96-10p and (d) L96-40p, for increasing levels of the assumed observation error, $\sigma_\mathrm{obs} = k\,\sigma_\epsilon$, where $\sigma_\epsilon$ is the standard deviation of the noise in the data and $k$ is given in the legends. Circles: every component observed; triangles: $y$ alone (L63) or every other site (Lorenz-96). Each marker is a run that does not fail, of $40$ windows of $210$ analyses (L63 and the ten-site ESNs) or ten windows of $48$ analyses (L96-40p, with the LETKF), with $m = 10$ members on L63 and L96-40p, and $64$ on the ten-site ESNs. Line: a calibrated ensemble, i.e., a spread equal to the error; the ensemble is over-dispersed above the line and under-dispersed below it. Grey band: a spread within a factor of two of the error, i.e., $e \in [s/2,\, 2s]$.}
  \label{fig:uncertainty}
\end{figure}

The ensemble quantifies the uncertainty of the state if its spread matches its error, i.e., if its spread--error ratio is close to one (\S\ref{sec:results}). Figure~\ref{fig:uncertainty} shows the spread against the error for increasing levels of the assumed observation error, $\sigma_\mathrm{obs}$, with every component observed and with partial observations. First, with every component observed and $\sigma_\mathrm{obs}$ equal to the noise in the data, the error is the smallest and the ensemble is the closest to calibrated, with ratios of $0.8$ to $1.0$. Second, a larger $\sigma_\mathrm{obs}$ increases both the error and the spread, because the analysis weighs the data less (\S\ref{sec:uq_method}). On L63, the spread grows faster than the error, and the ensemble becomes over-dispersed, with a ratio of $1.7$ at five times the noise, and of $2.2$ at ten times the noise. On the ten-site ESNs, the ensemble remains under-dispersed, with ratios of $0.6$ to $0.7$. On L96-40p, the ensemble is calibrated at five times the noise, with a ratio of $1.05$, and over-dispersed at the $\sigma_\mathrm{obs}$ of Table~\ref{tab:cases}, with a ratio of $2.0$. Third, with sparse data, a $\sigma_\mathrm{obs}$ equal to the noise is no longer the most accurate. On L96-40p from every other site, the ensemble collapses, with a ratio of $0.23$, and its error, $0.09$, exceeds that with the $\sigma_\mathrm{obs}$ of Table~\ref{tab:cases}, $0.08$. The $\sigma_\mathrm{obs}$ of Table~\ref{tab:cases}, therefore, trades accuracy with dense data for robustness with sparse data. The inflation is necessary on the ten-site system: without it, the spread of L96-10p drops from $0.76$ to $0.49$ times its error, and the filter fails in one of the ten windows. The ensemble of the forcing is under-dispersed, with a spread of $0.75$ times its error (median over the $63$ runs of \S\ref{sec:param_results}).

\section{Conclusions}
\label{sec:conclusions}

A trained echo state network (ESN) can forecast only once its reservoir state is synchronized with the system. With a washout, this synchronization requires the full state at every time step. To bypass the washout, we recast the deployment of the ESN as sequential state estimation. An ensemble square-root Kalman filter corrects the inputs and the reservoir states of an ensemble whose members start from states visited during training; the training is unchanged. 
We extend the method to (i) observations that leave a symmetry of the system unresolved, i.e., bimodal posteriors, via a Gaussian-mixture analysis; (ii) parametric ESNs, in which the physical parameter enters the readout, via joint state and parameter estimation; and (iii) parallel ESNs, via a localization.

We test the method on the Lorenz-63 system and on the Lorenz-96 system with 10 and 40 sites. First, with the full state observed, three analyses synchronize the ESN, i.e., $15\%$ of the data of the training washout, whereas a washout of as many data does not. Because a washout of a quarter of a Lyapunov time already forecasts as long as the filter, the advantage of the filter is the deployment from few data and from partial observations. Second, the filter deploys a single full-state ESN from a subset of its components, which a washout cannot do. On the Lorenz-63 system, the filter synchronizes the ESN from every subset except $z$ alone in at least $96\%$ of the initializations. From $z$ alone, which does not determine the sign of $x$ and $y$, the Gaussian-mixture analysis synchronizes the ESN in up to $91\%$ of the runs, against $16$ to $46\%$ with one Gaussian. On the ten-site Lorenz-96 system, direct replacement diverges from five, three and two sites, whereas the filter synchronizes the ESN in every window. Third, the filter estimates a forcing that is absent from the training data with a median error of $0.2\%$ inside the trained range and of at most $1.2\%$ beyond it, and the estimated forcing lengthens the valid forecast in chaos from at most $0.7$ to up to $7.6\,T_\lambda$. Because the forcing enters the readout rather than the input, $1$ in $825$ closed loops leaves the attractor, against $1$ in $7$ with the forcing in the input. Fourth, on the 40-site system with every site observed, localization reduces the number of members that the synchronization requires from $40$ to $5$. Fifth, the ensemble spread signals, without the true state, when the ESN synchronizes and when the filter fails. With the assumed observation error equal to the noise in the data, the ensemble is close to calibrated, with spread--error ratios of $0.8$ to $1.0$. A larger assumed error over-disperses the ensemble on the Lorenz-63 and 40-site systems, whereas the ten-site ensemble remains under-dispersed, which is consistent with a model error of the ESN that the ensemble does not represent.

Future work will focus on (i) representing the model error of the ESN in the ensemble to calibrate its spread; and (ii) deploying ESNs on experimental data. Because the training is unchanged, the method applies to any trained ESN. This opens the deployment of echo state networks and other recurrent architectures to systems that are measured partially, sparsely in time, and at unknown operating conditions.

\section*{Acknowledgements}
This work is supported by Schmidt Sciences, LLC.

\section*{Data and code availability}
The single ESNs are trained with the open-source Python package \texttt{echostatenetwork}. The scripts, configurations and trained ESNs of the experiments in this paper are available at \url{https://github.com/andreanovoa/romda} (release 4.0.0). 

\section*{Use of generative AI tools}
During the preparation of this work, the author used Claude Code to streamline the code and to proofread the manuscript. After using this tool/service, the author reviewed and edited the content as needed and takes full responsibility for the content of the published article.

\appendix
\counterwithin{figure}{section}
\counterwithin{table}{section}
\counterwithin{equation}{section}
\section{Trained ESNs and their verification}
\label{app:verification}

Table~\ref{tab:esns} lists the six ESNs, with the selected hyperparameters and the ranges of the search. The hyperparameters of the single ESNs are selected by Bayesian optimization on $32$ validation folds of $5\,T_\lambda$~\citep{racca_robust_2021}, and those of L96-40p on a grid of $60$ values by the median valid time (Eq.~\eqref{eq:success}) of $30$ validation forecasts. A candidate whose closed loop of $50\,T_\lambda$ leaves the attractor, or changes the standard deviation of an output by more than $20\%$, is penalized or, for L96-40p, discarded~\citep{platt_systematic_2022}. The single ESN L96-40s is trained after a washout of $100$ steps, against $20$ for the other ESNs. On L96-40, patches of five sites forecast for $2.9\,T_\lambda$ in validation, against $1.8$ to $2.2\,T_\lambda$ for patches of two, four and eight sites.

\begin{figure}
  \centering
  \includegraphics[width=\columnwidth]{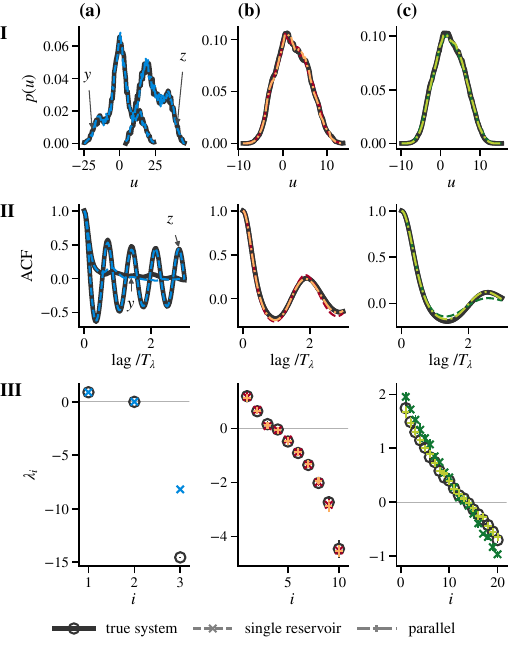}
  \caption{{Verification of the ESNs.} Closed-loop ESNs against the true systems. (a) L63. (b) L96-10s and L96-10p. (c) L96-40p and L96-40s. (I) Probability density over $400\,T_\lambda$. (II) Autocorrelation function (ACF). (III) Leading Lyapunov exponents over $50\,T_\lambda$, mean and standard deviation over ten starts. In (a), rows I and II show the components $y$ and $z$; in (b) and (c), the density and the ACF are over all sites, which are statistically equivalent.}
  \label{fig:verification}
\end{figure}

Every ESN reproduces the dynamics and the statistics of its system (Fig.~\ref{fig:verification}). After a washout of $2\,T_\lambda$ at the start of ten test windows, the forecast is valid for $2.2$ to $3.7$ Lyapunov times ($0.72\,T_\lambda$ for L96-40s). Over $400\,T_\lambda$ of closed loop, the density, standard deviation and autocorrelation of every component match those of the system, and every ESN stays on the attractor for $500\,T_\lambda$ from every start. The leading Lyapunov exponents~\citep{pathak_using_2017,margazoglou_stability_2023}, over $50\,T_\lambda$, match within their spread over the starts, except the leading exponent of L96-40s, $1.95$ against $1.74$, and the third exponent of L63, $-8.2$ against $-14.6$: a reservoir of $100$ units need not reproduce the strongly contracting direction.
\begin{table*}
  \centering
  \footnotesize
  \setlength{\tabcolsep}{4pt}
  \caption{The six ESNs. Training length in steps of the record, $\delta t$, and in $T_\lambda$; training washout in ESN steps, $\Delta t$ (Table~\ref{tab:cases}). The parametric ESN is trained on records of equal length at every forcing, given in steps only, because $T_\lambda$ is not finite for $F \le 4.5$. Hyperparameters: the selected value, with the range of the search after the name.}
  \label{tab:esns}
  \begin{tabular}{llcccccc}
    \toprule
     & & Lorenz-63 & \multicolumn{3}{c}{Lorenz-96, $N_u = 10$} & \multicolumn{2}{c}{Lorenz-96, $N_u = 40$} \\
    \cmidrule(lr){3-3}\cmidrule(lr){4-6}\cmidrule(lr){7-8}
     & & single & single & parallel & single, parametric & single & parallel \\
    Description & Symbol & L63 & L96-10s & L96-10p & parametric L96-10s & L96-40s & L96-40p \\
    \midrule
    Training regime & & $\rho=28$ & $F=8$ & $F=8$ & \begin{tabular}[t]{@{}c@{}}$F = \{3.5$, $3.75$, $4$, $4.25$,\\ $4.5$, $5$, $5.5$, $6$, $6.25$,\\ $6.75$, $7$, $7.5$, $8$\}\end{tabular} & $F=8$ & $F=8$ \\
    Training length (steps of $\delta t$; $T_\lambda$) & $T_\mathrm{train}$ & $5400$; $100$ & $33600$; $400$ & $33600$; $400$ & $3592$ per value of $F$ & $47200$; $800$ & $23600$; $400$ \\
    Reservoirs $\times$ units & & $1\times100$ & $1\times500$ & $2\times500$ & $1\times1000$ & $1\times6000$ & $8\times500$ \\
    Sites per patch; halo & & -- & -- & $5$; $2$ & -- & -- & $5$; $2$ \\
    Training washout (steps; $T_\lambda$) & $T_w$ & $20$; $0.74$ & $20$; $0.48$ & $20$; $0.48$ & $20$; $0.48$ & $100$; $3.3$ & $20$; $0.67$ \\
    \midrule
    Spectral radius, $[0.2, 1.2]$ & $\rho_\mathrm{s}$ & $0.50$ & $0.30$ & $0.30$ & $0.20$ & $0.21$ & $0.23$ \\
    Input scaling, $[0.03, 10]$\textsuperscript{a} & $\sigma_\mathrm{in}$ & $0.56$ & $0.54$ & $0.92$ & $0.56$ & $2.2$ & $0.62$ \\
    Tikhonov factor, $[10^{-10}, 10^{-4}]$\textsuperscript{b} & $\beta$ & $10^{-9}$ & $10^{-9}$ & $10^{-10}$ & $10^{-8}$, fixed & $10^{-4}$, fixed & $10^{-10}$ \\
    \bottomrule
    \multicolumn{8}{@{}l@{}}{\scriptsize \textsuperscript{a}Searched on a logarithmic scale. \textsuperscript{b}Searched over $\{10^{-10}, 10^{-9}, 10^{-7}, 10^{-6}, 10^{-4}\}$, except where fixed.}
  \end{tabular}
\end{table*}

\section{The parameter in the input or in the readout}
\label{app:readout_param}

A parametric ESN can read the parameter, here the forcing, $F$, in the input or in the readout. In the input~\citep{kong_machine_2021}, the readout is that of Eq.~\eqref{eq:esn_update}, and the reservoir state evolves as
\begin{equation*}
    \vect{r}(t_{n+1}) = \tanh\!\left(\matr{W}\vect{r}(t_n) + \matr{W}_\mathrm{in}\vect{u}_\mathrm{in}(t_n) + \vect{w}^{F}_{\mathrm{in}}\tilde F + \vect{b}_\mathrm{in}\right),
\end{equation*}
where $\tilde F$ is the normalized forcing, and $\vect{w}^{F}_{\mathrm{in}} \in \mathbb{R}^{N_r}$ is its input weight. In the readout (\S\ref{sec:param_method}), the reservoir state evolves as in Eq.~\eqref{eq:reservoir}, and the readout is
\begin{equation*}
    \vect{u}(t_{n+1}) = \matr{W}_\mathrm{out}[\vect{r}(t_{n+1});\ 1] + \vect{w}^{F}_{\mathrm{out}}\tilde F,
\end{equation*}
where $\matr{W}_\mathrm{out}$ and the output weight of the forcing, $\vect{w}^{F}_{\mathrm{out}} \in \mathbb{R}^{N_u}$, are trained together by the ridge regression of Eq.~\eqref{eq:ridge} on the records of every training forcing. In open loop, the reservoir state does not depend on the forcing. In closed loop, the forcing reaches the reservoir through the feedback, $\matr{W}_\mathrm{in}\vect{w}^{F}_{\mathrm{out}}\tilde F$. The two choices are complementary. With the forcing in the input, the reservoir state depends nonlinearly on $F$, so one reservoir can hold a different attractor at each forcing. However, the output is sensitive to the forcing at a fixed reservoir state, so each member must start from a washout at the forcing that it carries. With the forcing in the readout, the washout needs no value of $F$. The reservoir state still encodes the regime, because it is driven by data generated at that forcing, so the filter updates $F$ through its covariance with the reservoir state.

We compare the parametric L96-10s of \S\ref{sec:param_results} with a single ESN of the same size, records and hyperparameter search that reads the forcing in the input, with an input scaling of the forcing of $0.01$.  Each ESN runs $2475$ closed loops of $400$ time units at $165$ forcings from $0.06$ to $10$, after washouts of $2\,T_\lambda$, five each on the training record of the nearest forcing, on random training records and on the true trajectory. A loop leaves the attractor when $|\vect{u}|$ exceeds three times the largest $|\vect{u}|$ of the training records.
At $\beta=10^{-8}$, the two ESNs forecast for comparable times, $2.1$ to $4.7\,T_\lambda$ at $F = 6.5$ to $8.5$, and estimate the forcing with comparable accuracy, with median errors of $0.32\%$ (input) and $0.28\%$ (readout). They differ in their sensitivity to the forcing. At a fixed reservoir state, one step changes the output by $2.6$ per unit of $F$ with the forcing in the input, against $0.060$ in the readout, and $1$ in $7$ loops leaves the attractor ($1$ in $23$ inside the trained range), against $1$ in $825$, all at $F \geq 9.7$. At $\beta = 10^{-12}$, the sensitivity in the input grows to $44$, more than half of the loops leave the attractor ($2$ in $5$ inside the trained range), and the filter estimates $F = 8.5$ with an error of $4.0\%$, whereas in the readout the sensitivity stays at $0.063$, no loop leaves the attractor, and the error is $0.5\%$.

\section{Localized correlations}
\label{app:covariance}

\begin{figure}
  \centering
  \includegraphics[width=.9\columnwidth]{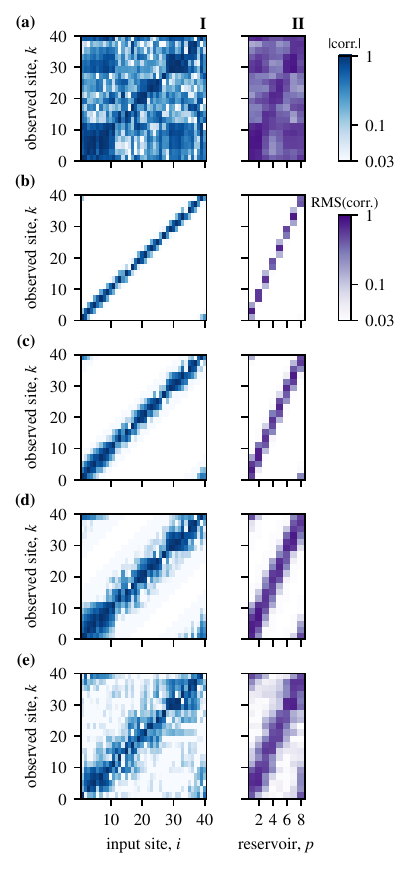}
  \caption{Forecast correlations between the state and the predicted observables on L96-40p (every other site observed, last analysis of one window), on a logarithmic colour scale from $0.03$ to $1$. (I) Input sites, magnitude. (II) Reservoirs, root mean square over the units. (a) $m = 10$ members; (b--d) the same, tapered with a half-width of two (the filter), four and eight sites; (e) $m = 1000$ members.}
  \label{fig:localized_covariance}
\end{figure}

Figure~\ref{fig:localized_covariance} shows why the global filter needs many members on L96-40p. With $10$ members, the sampled correlation between an input site and an observed site is as large far from the site as near it (Fig.~\ref{fig:localized_covariance}a), whereas the correlations of $1000$ members decay within a few sites of the diagonal (Fig.~\ref{fig:localized_covariance}e). The Gaspari--Cohn taper with the half-width of the filter, two sites, removes the spurious correlations and keeps those near the diagonal (Fig.~\ref{fig:localized_covariance}b). Wider tapers, of four and eight sites, keep more of the spurious correlations of the small ensemble (Fig.~\ref{fig:localized_covariance}c--d).

\section{Gaussian-mixture analysis}
\label{app:mixture}

At each analysis, the Gaussian-mixture filter of \S\ref{sec:mixture_method} (i) fits a mixture of up to two Gaussians, with one diagonal covariance, to the forecast inputs, $\vect{u}^{\rm f}_j$, scaled by the standard deviation of each component of the training data, by expectation maximization~\citep{dempster_maximum_1977}, and assigns each member to its most likely Gaussian, $k$; (ii) inflates the $m_k$ members of each Gaussian about their own mean, Eq.~\eqref{eq:inflate}, and updates them with Eq.~\eqref{eq:ensrkf}, with the mean and the deviations of these $m_k$ members; (iii) updates the weight of each Gaussian with the likelihood of the data~\citep{alspach_nonlinear_1972}
\begin{equation}
    w^{\rm a}_k \propto \frac{m_k}{m}\,\mathcal{N}\big(\vect{d};\ \bar{\vect{q}}^{\rm f}_k,\ \matr{C}_{dd} + \matr{C}^{\rm f}_{qq,k}\big),
    \label{eq:mixture_weights}
\end{equation}
where $\bar{\vect{q}}^{\rm f}_k$ and $\matr{C}^{\rm f}_{qq,k}$ are the mean and the covariance of the predicted observables of Gaussian $k$, and the weights sum to one; and (iv) reallocates the members in proportion to $w^{\rm a}_k$. A Gaussian that loses members drops them at random, and a Gaussian that gains members draws them from its analysis, $\bar{\vect{\psi}}^{\rm a}_k + \matr{\Psi}'^{\rm a}_k\vect{\xi}/\sqrt{m_k-1}$, with $\vect{\xi}$ drawn from $\mathcal{N}(\vect{0}, \mathbb{I}_{m_k})$. With one Gaussian, the filter is the EnSRKF of Eq.~\eqref{eq:ensrkf_update}.

The filter keeps a second Gaussian only if each Gaussian has at least three members, and if the second Gaussian lowers the Bayesian information criterion~\citep{schwarz_estimating_1978} by more than $10$, i.e., very strong evidence~\citep{kass_bayes_1995}. Without this margin, the criterion splits $2$ to $20\%$ of samples of $10$ to $30$ members drawn from one three-dimensional Gaussian; with the margin, $1\%$ or fewer (not shown). The scaling by the training data, instead of the spread of the ensemble, keeps a synchronized ensemble in one Gaussian: the variance of a Gaussian is at least $10^{-2}$ in these units, so an ensemble narrower than a tenth of the standard deviation of the training data is not split.

\bibliographystyle{cas-model2-names}
\bibliography{cas-refs}

\end{document}